\documentclass[aps,prc,showpacs,twocolumn]{revtex4}
\usepackage{graphicx}
\usepackage{dcolumn}
\usepackage{bm}
\usepackage{CJK}
\usepackage{color}
\usepackage{amsfonts}
\usepackage{mathrsfs}
\usepackage{amssymb}
\usepackage{amsmath}

\usepackage[colorlinks,linkcolor=blue,anchorcolor=blue,citecolor=blue]{hyperref}

\newcommand{\beq}{\begin{equation}}
\newcommand{\eeq}{\end{equation}}
\newcommand{\beqn}{\begin{eqnarray}}
\newcommand{\eeqn}{\end{eqnarray}}
\newcommand{\bsub}{\begin{subequations}}
\newcommand{\esub}{\end{subequations}}

\newcommand{\br}{{\mathbf{r}}}

\begin{document}


 \title{Rapid structural change in low-lying states of neutron-rich Sr and Zr isotopes}

\author{H. Mei$^{1,2}$, J. Xiang$^1$, J. M. Yao$^{2,1}$, Z. P. Li$^1$, J. Meng$^{3,4,5}$ }

\affiliation{$^1$School of Physical Science and Technology, Southwest
University, Chongqing 400715, China}
\affiliation{$^2$Physique Nucl\'eaire Th\'eorique,
             Universit\'e Libre de Bruxelles, C.P. 229, B-1050 Bruxelles,
             Belgium}
\affiliation{$^3$School of Physics, Peking University, Beijing 100871, China}
\affiliation{$^4$School of Physics and Nuclear Energy Engineering, Beihang University, Beijing 100191, China}
\affiliation{$^5$Department of Physics, University of Stellenbosch, Stellenbosch, South Africa}

 \begin{abstract}
 The rapid structural change in low-lying collective excitation states of neutron-rich Sr and Zr isotopes is studied by solving a five-dimensional collective Hamiltonian with parameters determined by both relativistic mean-field and non-relativistic Skyrme-Hartree-Fock calculations using the PC-PK1 and SLy4 forces respectively. Pair correlations are treated in BCS method with either a separable pairing force or a density-dependent zero-range force. The isotope shifts, excitation energies, electric monopole and quadrupole transition strengths are calculated and compared with corresponding experimental data. The calculated results with both the PC-PK1 and SLy4 forces exhibit a picture of spherical-oblate-prolate shape transition in neutron-rich Sr and Zr isotopes. Compared with the experimental data, the PC-PK1 (or SLy4) force predicts a more moderate (or dramatic) change in most of the collective properties around $N=60$. The underlying microscopic mechanism responsible for the rapid transition is discussed.

 \end{abstract}
 \date{\today}
\maketitle


 \section{Introduction}
The spectroscopy of nuclear low-lying states provides rich information about the interplay of nuclear collectivity and single-particle structure. For neutron-rich Sr and Zr isotopes, lots of spectroscopic data have been measured~\cite{Jung80,Schussler80,Kawade82,Eberth88,Mach89,Mach91,
Lhersonneau94,Urban01,Hager06,Goodin07,Charlwood09}. The abrupt changing in the lifetimes and excitation energies of $2^+_1$ states, two-neutron separation energies and rms charge radii indicates a sudden onset of large quadrupole deformation at neutron number $N=60$. In the past decades, various models have been employed to study such a dramatic transition in the low-energy structure of these nuclei ~\cite{Federman78,Kumar85,Galeriu86,Michiaki90,Moller95,Skalski97,Lalazissis99,
Holt00,Xu02,Ramos05,Geng05,Lalkovski09,Isacker10,Boyukata10}, the description of which has been a challenge in theoretical nuclear physics.

Two main mechanisms based on shell models and mean-field approaches were proposed to explain the sudden onset of large nuclear collectivity at $N=60$, namely, a strong isoscalar proton-neutron interaction between particles occupying the $g_{9/2}$-$g_{7/2}$ spin-orbit partners~\cite{Federman78}, and the occupation of low-$K$ components of the $h_{11/2}$ neutron orbit~\cite{Bonche85,Werner94} respectively. Besides these two factors, the simultaneous polarization of $2p_{1/2}, 2p_{3/2}$, and $1f_{5/2}$ proton orbits as one goes from $^{96}$Sr
to $^{98}$Sr and from $^{98}$Zr to $^{100}$Zr was pointed out to be the the major
factor in the more recent projected shell model study~\cite{Verma08}. However, in most of the previous studies for nuclear low-lying states with, for instance the (projected) shell model, either a phenomenological effective interaction or effective charges for neutrons and protons defined within a specific valence space was introduced. In particular, it has been shown recently that the large-scale shell model calculations with a more extended model space and carefully adjusted effective interaction are able to reproduce the low-spin spectroscopic data of $^{90-98}$Zr, but fail to give large deformation properties for $^{100}$Zr~\cite{Sieja09}. The mechanism responsible  for the rapid transition in the structure of low-lying states in Sr and Zr isotopes around $N=60$ requires further investigation.

Nuclear energy density functional theory (DFT) is nowadays one of the most
important microscopic approaches for large-scale nuclear structure calculations in medium and
heavy nuclei. The main ingredient of DFT is the energy density functional (EDF) that depends on densities and currents representing distributions of nucleonic matter, spins, momentum, and kinetic energy and their derivatives. In the past decades, a lot of efforts have been devoted into finding out an EDF with reliable and good predictions by optimizing about 10 universal parameters to basic properties of nuclear matter and some selected nuclei. At present, there are three types of most successful EDFs, i.e., the non-relativistic Skyrme and Gogny forces and the effective relativistic Lagrangian, being employed extensively in the description of nuclear structure properties~\cite{Lalazissis04}.

The nuclear DFT of single-reference state (SR-DFT) with constraint on quadrupole moments, in the context of Hartree-Fock (HF) or Hartree-Fock-Bogoliubov (HFB) methods, has already been adopted to study the evolution of deformation energy surfaces in $\beta$-$\gamma$ plane (see the recent work~\cite{Lu11}) for the neutron-rich Sr and Zr isotopes~\cite{Bonche85,Guzman10,Xiang12}. Even though, all of these studies have indeed shown the increasing of deformations based on the shift of global minimum. However, most of these studies are focused on the properties of nuclear mean-field ground states. In particular, a competing weakly oblate deformed minima coexisting with a large prolate one has been found in nuclei around $N=60$, in which case the dynamic correlation effects from quadrupole fluctuation are expected to be significant. Moreover, the evolution behavior of low-lying excited states is sensitive to the balance between the oblate and prolate minima. Therefore, it is necessary to extend these studies for the low-lying excited states in Sr and Zr isotopes.

In the context of generator-coordinate method (GCM), the framework of SR-DFT has been extended for studying the nuclear low-lying states. In Ref.~\cite{Skalski93}, the collective quadrupole and octupole excitations in Zr isotopes were studied using a basis generated by the HF+BCS calculations with the SkM* effective interaction. However, the low-lying states have not been described quite well, probably due to the missing of effects from triaxiality and angular momentum projections. In recent years, the GCM has been extended much further by implementing the exact one-dimensional~\cite{Valor00,Guzman02,Niksic06} or three-dimensional~\cite{Bender08,Yao09,Yao10,Rodriguez10} angular momentum projection or together with particle number projection before or after variation for the mean-field states in the modern EDF calculations. The dynamic correlation effects related to the symmetry restoration and quadrupole fluctuation (along both $\beta$ and $\gamma$ directions) around the mean-field minimum are included naturally without introducing any additional parameters. However, the application of these methods with triaxiality for systematic study is still much time-consuming. Up to now, such kind of study is mostly restricted to light nuclei~\cite{Yao11-Mg,Yao11-C} and some specific medium heavy nuclei~\cite{Tomas11PRC,Tomas11PLB}.

As a Gaussian overlap approximation of GCM, the collective Hamiltonian with parameters determined by self-consistent mean-field calculations is much simple in numerical calculations, and has achieved great success in description of nuclear low-lying states~\cite{Libert99,Prochniak04,Niksic09,Li09,Niksic11} and impurity effect of $\Lambda$ hyperon in nuclear collective excitation~\cite{Yao11-lambda}. In particular, a systematic study of low-lying states for a large set of even-even nuclei has been carried out with the Gogny D1S force mapped collective Hamiltonian and good overall agreement with the low-lying spectroscopic data has been achieved~\cite{Delaroche10}. However, some fine structures along the isotonic or isotopic chains are still not reproduced satisfactorily. For Sr and Zr isotopes, the evolution of isotope shifts and the change in the properties of low-lying states are found to be more moderate in comparison with the data. Therefore, in this work, we will examine the evolution of low-lying states obtained from the calculations with two other popular EDFs, i.e., the non-relativistic Skyrme force and the effective relativistic Lagrangian. The difference in the results of these two calculations will be emphasized.

The paper is organized as follows. In Sec.~\ref{Sec.II} we will
introduce briefly our method used to study the low-lying states in neutron-rich Sr and Zr isotopes. The results and discussions will be presented in Sec.\ref{Sec.III}. A summary is made in Sec.~\ref{Sec.IV}.

 \section{The method}%
 \label{Sec.II}
The quantized five-dimensional collective Hamiltonian (5DCH) that describes the nuclear excitations of quadrupole vibrations, rotations, and their couplings can be
written in the form~\cite{Libert99,Prochniak04,Li09}
\begin{equation}
\label{hamiltonian-quant}
\hat{H} =
\hat{T}_{\textnormal{vib}}+\hat{T}_{\textnormal{rot}}
              +V_{\textnormal{coll}} \; ,
\end{equation}
where $V_{\textnormal{coll}}$ is the collective potential. The
vibrational kinetic energy reads,
\begin{eqnarray}
\hat{T}_{\textnormal{vib}}
 &=&-\frac{\hbar^2}{2\sqrt{wr}}
   \left\{\frac{1}{\beta^4}
   \left[\frac{\partial}{\partial\beta}\sqrt{\frac{r}{w}}\beta^4
   B_{\gamma\gamma} \frac{\partial}{\partial\beta}\right.\right.\nonumber\\
  && \left.\left.- \frac{\partial}{\partial\beta}\sqrt{\frac{r}{w}}\beta^3
   B_{\beta\gamma}\frac{\partial}{\partial\gamma}
   \right]+\frac{1}{\beta\sin{3\gamma}} \left[
   -\frac{\partial}{\partial\gamma} \right.\right.\nonumber\\
  && \left.\left.\sqrt{\frac{r}{w}}\sin{3\gamma}
      B_{\beta \gamma}\frac{\partial}{\partial\beta}
    +\frac{1}{\beta}\frac{\partial}{\partial\gamma} \sqrt{\frac{r}{w}}\sin{3\gamma}
      B_{\beta \beta}\frac{\partial}{\partial\gamma}
   \right]\right\},
 \end{eqnarray}
and rotational kinetic energy,
\begin{equation}
\hat{T}_{\textnormal{\textnormal{\textnormal{rot}}}} =
\frac{1}{2}\sum_{k=1}^3{\frac{\hat{J}^2_k}{\mathcal{I}_k}},
\end{equation}
with $\hat{J}_k$ denoting the components of the angular momentum in
the body-fixed frame of a nucleus. It is noted that the mass
parameters $B_{\beta\beta}$, $B_{\beta\gamma}$, $B_{\gamma\gamma}$,
as well as the moments of inertia $\mathcal{I}_k$, depend on the
quadrupole deformation variables $\beta$ and $\gamma$,
\begin{equation}
\mathcal{I}_k = 4B_k\beta^2\sin^2(\gamma-2k\pi/3),~~ k=1, 2, 3.
\end{equation}
Two additional quantities that appear in the expression for the
vibrational energy: $r=B_1B_2B_3$, and
$w=B_{\beta\beta}B_{\gamma\gamma}-B_{\beta\gamma}^2 $, determine the
volume element in the collective space. The corresponding eigenvalue
problem is solved using an expansion of eigenfunctions in terms of a
complete set of basis functions that depend on the deformation
variables $\beta$ and $\gamma$, and the Euler angles $\phi$,
$\theta$ and $\psi$~\cite{Pro.99}.

The dynamics of the 5DCH is governed by the seven
functions of the intrinsic deformations $\beta$ and $\gamma$: the
collective potential $V_{\rm coll}$, the three mass parameters:
$B_{\beta\beta}$, $B_{\beta\gamma}$, $B_{\gamma\gamma}$, and the
three moments of inertia $\mathcal{I}_k$. These functions are determined
by the relativistic mean-field (RMF)+BCS calculations using the PC-PK1 force~\cite{Zhao10} for the particle-hole ($ph$) and the separable pairing force (adjusted to reproduce the pairing
properties of the Gogny force D1S in nuclear matter)~\cite{Tian09,Niksic10}  for the particle-particle ($pp$) channels or by the Skyrme-Hartree-Fock (SHF)+BCS calculations using the SLy4 force~\cite{Chabanat98} for the $ph$ channel and a density-dependent $\delta$-force in the $pp$ channel,
 \begin{equation}
 V (\br_1,\br_2) = V^{pp}_0 \left[ 1- \dfrac{\rho(\br)}{\rho_0}\right] \delta(\br_1-\br_2),
 \end{equation}
with a strength of $V^{pp}_0=-1000$~MeV fm$^3$ and $\rho_0=0.16$ fm$^{-3}$ for
both neutrons and protons and with a soft cutoff at 5~MeV above and below
the Fermi energy as defined in Ref.~\cite{Rig99}. A constraint on the deformation parameters of both $\beta$, ranging from $0$ to $0.8$ ($\Delta\beta=0.05$) and $\gamma$, ranging from $0^\circ$ to $60^\circ$ ($\Delta\gamma=6^\circ$) is imposed in both calculations.

 \begin{figure*}[]
  \centering
 \includegraphics[width=7cm]{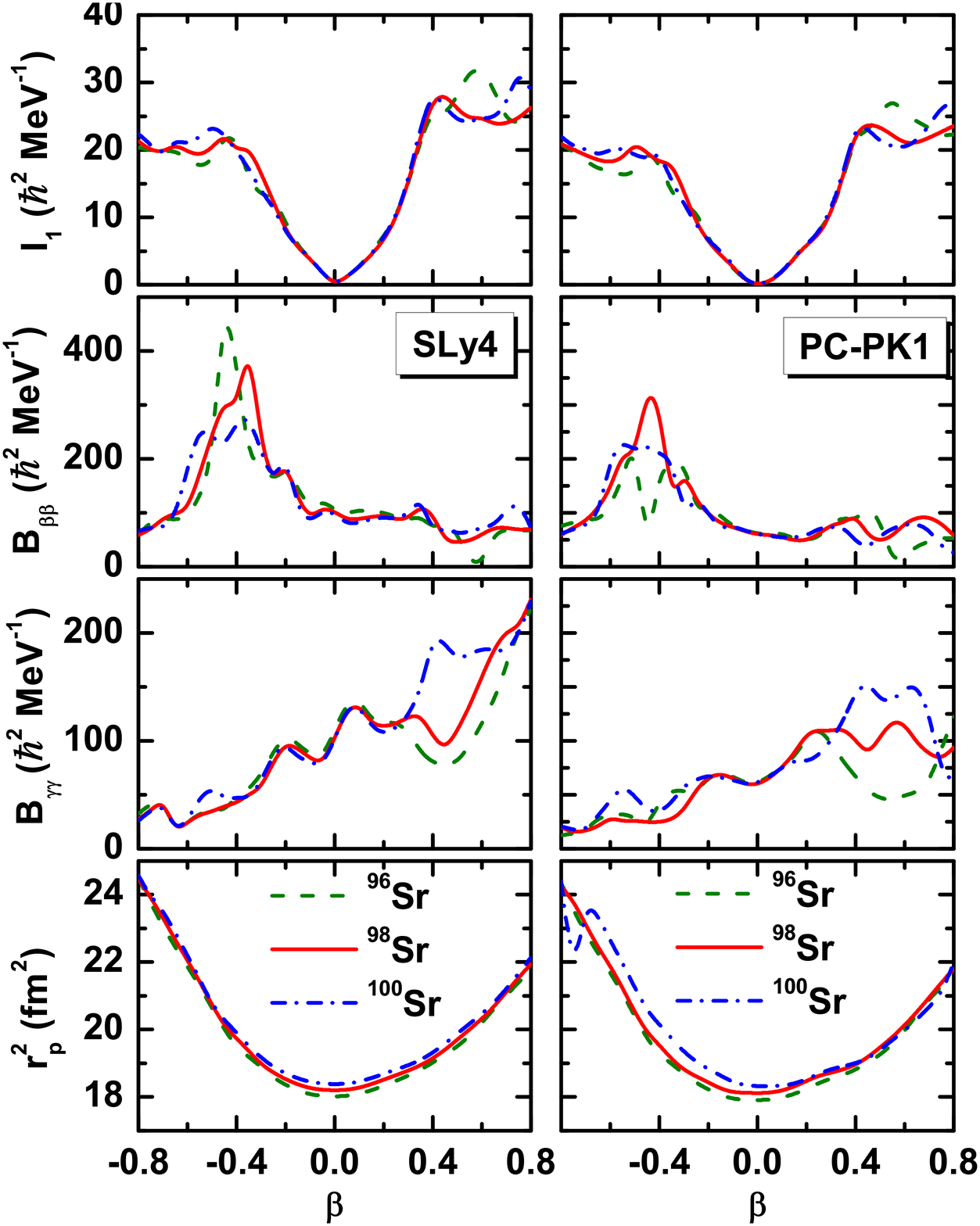}
 \includegraphics[width=7cm]{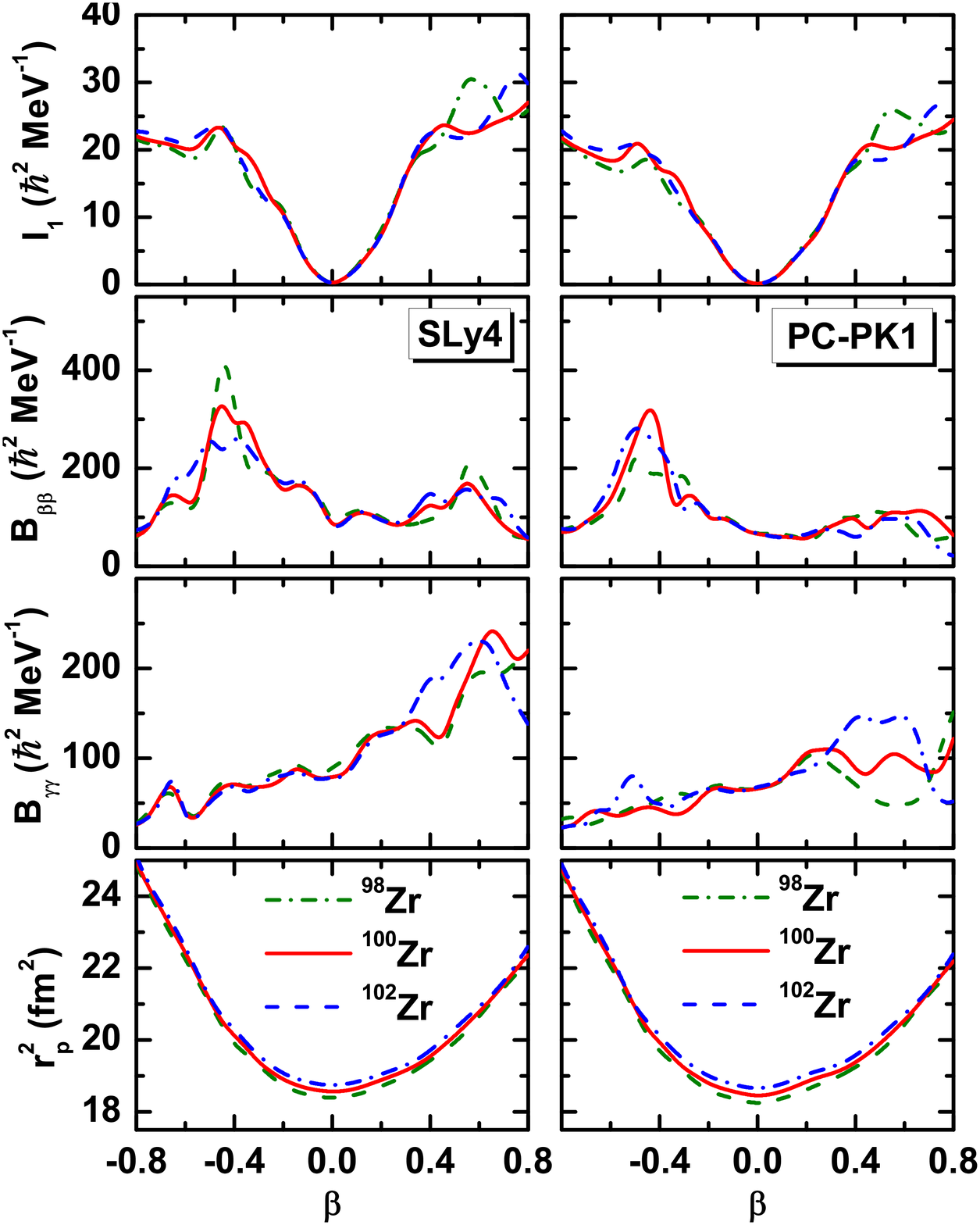}
 \caption{(Color online) Moment of inertia along $x$-direction $I_1$, mass parameters $B_{\beta\beta}, B_{\gamma\gamma}$ and squared proton radii $r^2_p$ in 5DCH for (left panel) $^{96,98,100}$Sr and (right panel) $^{98,100,102}$Zr as functions of axial deformation parameter $\beta$ determined from the mean-field calculations with both the SLy4 and PC-PK1 forces.}
 \label{Parameters}
\end{figure*}

The moments of inertia $\mathcal{I}_\kappa$ are calculated using the Inglis-Belyaev formula~\cite{Ing.56,Bel.61}
\begin{equation}
\label{Inglis-Belyaev}
 \mathcal{I}_{k} =
 \sum_{i,j}{\frac{\left(u_iv_j-v_iu_j \right)^2}{E_i+E_j}
  \langle i |\hat{J}_{k} | j  \rangle |^2},
\end{equation}
where $k=1, 2, 3$ denotes the axis of rotation, and the
summation of $i, j$ runs over the proton and neutron quasiparticle
states. The mass parameters $B_{\mu\nu}(\beta,\gamma)$ are given by~\cite{GG.79}
\begin{equation}
\label{masspar-B} B_{\mu\nu}(\beta,\gamma)=\frac{\hbar^2}{2}
 \left[\mathcal{M}_{(1)}^{-1} \mathcal{M}_{(3)} \mathcal{M}_{(1)}^{-1}\right]_{\mu\nu}\;,
\end{equation}
with
\begin{equation}
 \label{MMatrix}
 \mathcal{M}_{(n),\mu\nu}(\beta,\gamma)=\sum_{i,j}
 {\frac{\left\langle i\right|\hat{Q}_{2\mu}\left| j\right\rangle
 \left\langle j\right|\hat{Q}_{2\nu}\left| i\right\rangle}
 {(E_i+E_j)^n}\left(u_i v_j+ v_i u_j \right)^2}.
\end{equation}
 The mass parameters $B_{\mu\nu}$ in Eq.(\ref{masspar-B}) can be converted into the
 forms of $B_{\beta\beta}, B_{\beta\gamma}, B_{\gamma\gamma}$ by using
 the following relationships~\cite{Pro.99},
 \begin{equation}
 \begin{pmatrix}
 B_{\beta\beta} \\
 B_{\beta\gamma} \\
 B_{\gamma\gamma}  \\
 \end{pmatrix}
 =
 \begin{pmatrix}
   \cos^2\gamma & \sin2\gamma &  \sin^2\gamma \\
  -\dfrac{1}{2}\sin2\gamma   & \cos2\gamma & \dfrac{1}{2}\sin^2\gamma  \\
   \sin^2\gamma  & -\sin2\gamma & \cos^2\gamma  \\
 \end{pmatrix}
 \begin{pmatrix}
 B_{00}a_{00}\\
 B_{02}a_{02} \\
 B_{22}a_{22}  \\
 \end{pmatrix},
 \end{equation}
 where the coefficients $a_{02}=a_{00}/\sqrt a_{22}=a_{00}/2$, with
 $a_{00}=9r^4_0A^{10/3}/16\pi^2$, and $r_0=1.2$.

The potential $V_{\rm coll}$ in Eq.(\ref{hamiltonian-quant}) is given by,
\begin{equation}
\label{Vcoll} {V}_{\textnormal{coll}}(\beta,\gamma)
 = E_{\textnormal{tot}}(\beta,\gamma)
  - \Delta V_{\textnormal{vib}}(\beta,\gamma) - \Delta
  V_{\textnormal{rot}}(\beta,\gamma),
\end{equation}
where $E_{\textnormal{tot}}$ is the total energy from constrained mean-field calculations. The $\Delta V_{\textnormal{vib}}$ and $\Delta V_{\textnormal{rot}}$ are the zero-point-energy of vibration and rotation respectively,
 \beqn
 \Delta V_{\textnormal{vib}}(\beta,\gamma)
 &=&
 \frac{1}{4}{\rm Tr}[{\cal M}^{-1}_{(3)}{\cal M}_{(2)}],\\
 \Delta V_{\textnormal{rot}}(\beta,\gamma)
 &=& \sum_{\mu=-2,-1,1} \Delta V_{\mu\mu}(\beta,\gamma),
\eeqn
where the matrix ${\cal M}_{(n)}$ is determined by
Eq.(\ref{MMatrix}) with indices $\mu,\nu$ running over $0$ and $2$ and corresponding mass quadrupole operators defined as $\hat{Q}_{20}\equiv2z^2-x^2-y^2$  and $\hat{Q}_{22}\equiv x^2-y^2$. Moreover, $\Delta V_{\mu\mu}(\beta,\gamma)$ is calculated by
 \begin{equation}
 \Delta V_{\mu\nu}(\beta,\gamma)
 = \frac{1}{4} \frac{{\cal M}_{(2),\mu\nu}(\beta,\gamma)}{{\cal
 M}_{(3),\mu\nu}(\beta,\gamma)},
 \end{equation}
where ${\cal M}_{(n),\mu\nu}(\beta,\gamma)$ is also determined by
Eq.(\ref{MMatrix}), but with the intrinsic components of quadrupole
operator defined as,
 \begin{eqnarray}
  \hat Q_{2\mu}
 \equiv\left\{
 \begin{array}{cc}
 -2iyz, &  \mu=1  \\
 -2xz, &  \mu=-1 \\
  2ixy, &  \mu=-2 \\
 \end{array}
 \right.
 \end{eqnarray}

The details about the solution of constrained RMF+BCS and SHF+BCS equations have been given in Refs.~\cite{Xiang12} and~\cite{Bonche05} respectively.

 \section{Results and discussion}
 \label{Sec.III}

 \begin{figure*}[]
  \centering
 \includegraphics[width=7cm]{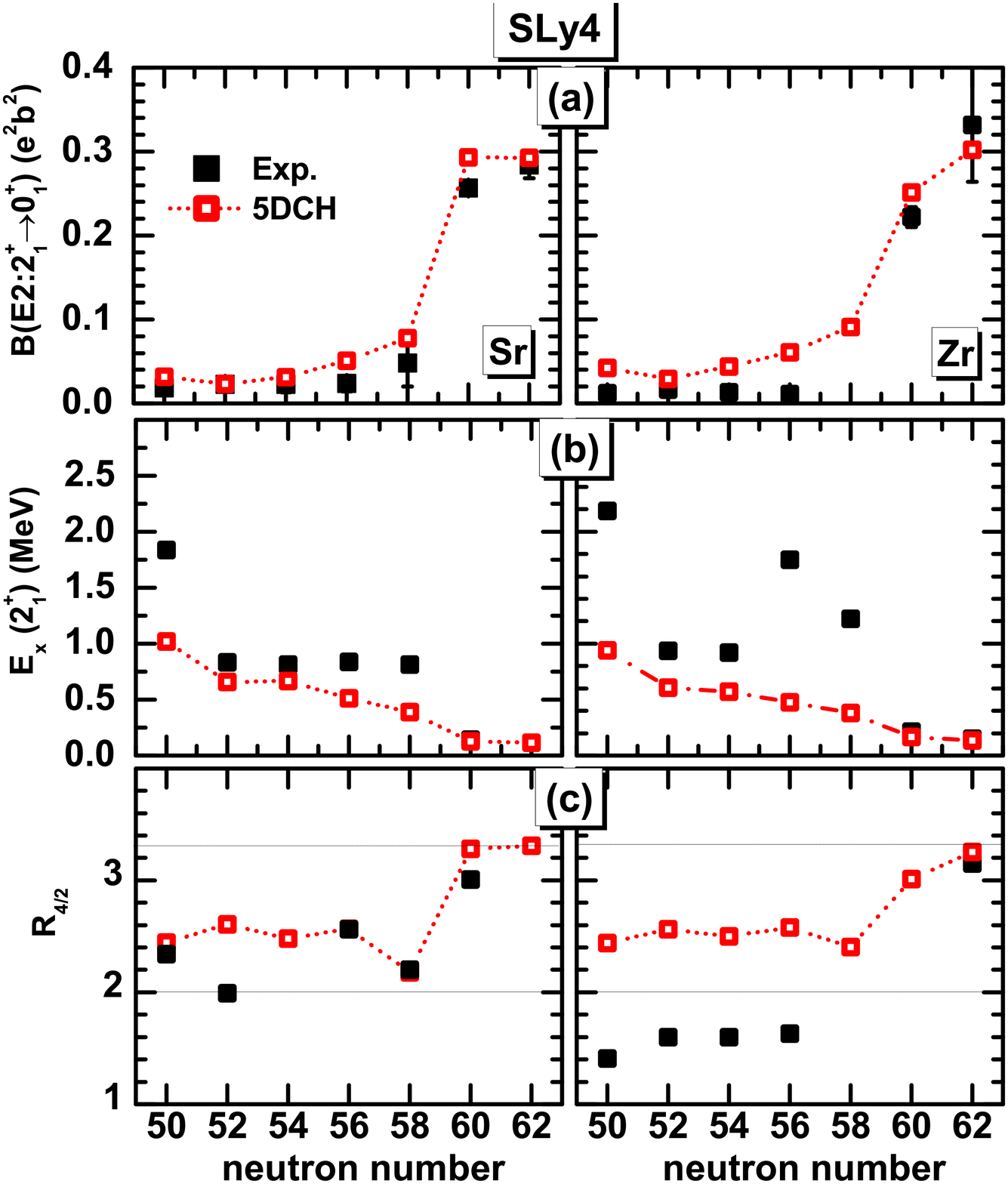}
 \includegraphics[width=7cm]{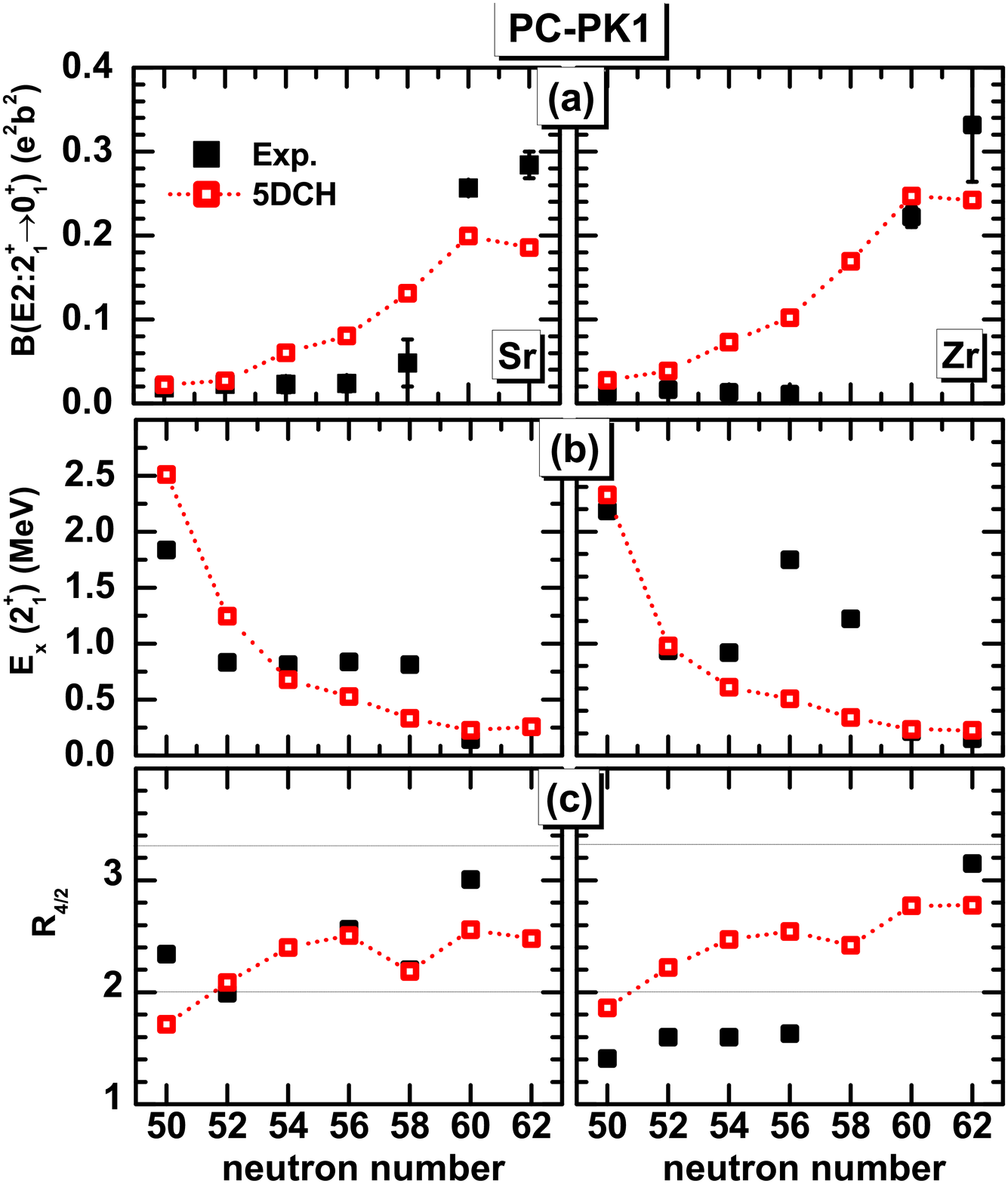}
 \caption{(Color online)  The electric quadrupole transition strength $B(E2:2^+_1\rightarrow 0^+_1)$,  excitation energy $E_x(2^+_1)$ and the ratio $R_{4/2} (\equiv E_x(4^+_1)/E_x(2^+_1))$ as functions of neutron number in Sr and Zr isotopes from the 5DCH calculations with (left panel) the SLy4 force and (right panel) the PC-PK1 force, in comparison with the experimental data~\cite{NNDC}. The $R_{4/2}$ values for vibration ($2.00$) and rotation ($3.33$) limits are indicated with horizontal dotted lines.}
 \label{SLy4:be2}
\end{figure*}

 \subsection{Collective parameters in Hamiltonian}

 Figure~\ref{Parameters} displays the moment of inertia along $x$-direction, mass parameters and squared proton radii in 5DCH for $^{96,98,100}$Sr and $^{98,100,102}$Zr isotopes as functions of axial deformation parameter $\beta$ determined from the mean-field calculations with both SLy4 and PC-PK1 forces. We note that the collective parameters in 5DCH do not change much in $^{96,98,100}$Sr and $^{98,100,102}$Zr isotopes at all the $\beta$-$\gamma$ deformation regions, except for the $B_{\beta\beta}$ with $\beta\simeq0.4$ and $\gamma=60^\circ$ as well as the $B_{\gamma\gamma}$ with $\beta\simeq0.5$ and $\gamma=0^\circ$. Moreover, the parameters in 5DCH determined from both SLy4 and PC-PK1 mean-field calculations are quite similar. For the mean squared proton radii $r^2_p$, the SLy4 predicts slightly larger value than the PC-PK1 by $\sim0.5\%$ systematically.

 \subsection{Spectroscopy of nuclear low-lying states}

 Figure~\ref{SLy4:be2} displays the electric quadrupole transition strength $B(E2:2^+_1\rightarrow 0^+_1)$,  excitation energy $E_x(2^+_1)$ of $2^+_1$ state and the ratio $R_{4/2} (\equiv E_x(4^+_1)/E_x(2^+_1))$ as functions of neutron number in Sr and Zr isotopes from the 5DCH calculations with both the SLy4 and the PC-PK1 forces. Systematically, the $B(E2:2^+_1\rightarrow 0^+_1)$ (or $E_x(2^+_1)$) values from the 5DCH calculations with both forces are increasing (or decreasing) with the neutron number up to $N=62$ in both Sr and Zr isotopes, which indicates the existence of transition from spherical to prolate deformed shapes. Compared with the experimental data, the evolution of $B(E2:2^+_1\rightarrow 0^+_1)$ and $R_{4/2}$ with respect to the neutron number around $N=60$ is much more dramatic (moderate) in the results of SLy4 (PC-PK1) calculations. Similar evolution behavior as that by the PC-PK1 force is also observed in the 5DCH calculations with the Gogny force D1S~\cite{Delaroche10}. However, this dramatic decreasing in $E_x(2^+_1)$ from $N=58$ to $N=60$ in Sr and Zr isotopes, together with the sudden increasing of $E_x(2^+_1)$ in $^{96}$Zr is not reproduced in both calculations.

 \subsection{Deformation energy surfaces and collective wave functions}
\begin{figure}[t]
  \centering
 \includegraphics[width=8.2cm]{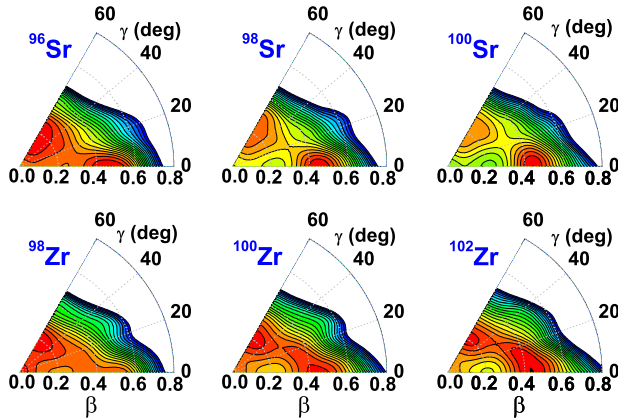}
 \caption{(Color online)The deformation energy surfaces of
$^{96,98,100}$Sr and $^{98,100,102}$Zr isotopes in the $\beta$-$\gamma$ plane, from the constrained SHF+BCS calculations with the SLy4 force for the $ph$ channel and density-dependent $\delta$ force for the $pp$ channel. All energies are normalized to the absolute minimum. Each contour line is separated by 0.5~MeV.}
 \label{Zr_Sr_SHF}
\end{figure}

\begin{figure}[t]
  \centering
 \includegraphics[width=8.2cm]{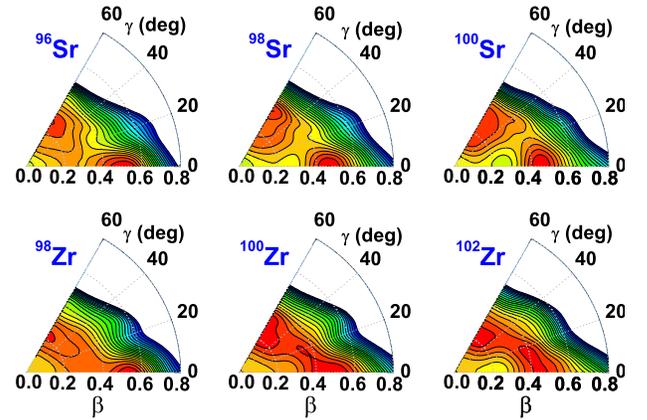}
 \caption{(Color online) Same as Fig.~\ref{Zr_Sr_SHF}, but from the constrained RMF+BCS calculations with the PC-PK1 force for the $ph$ channel and a separable force for the $pp$ channel.}
 \label{Zr_Sr_RMF}
\end{figure}

\begin{figure}[htp]
  \centering
 \includegraphics[width=8.2cm]{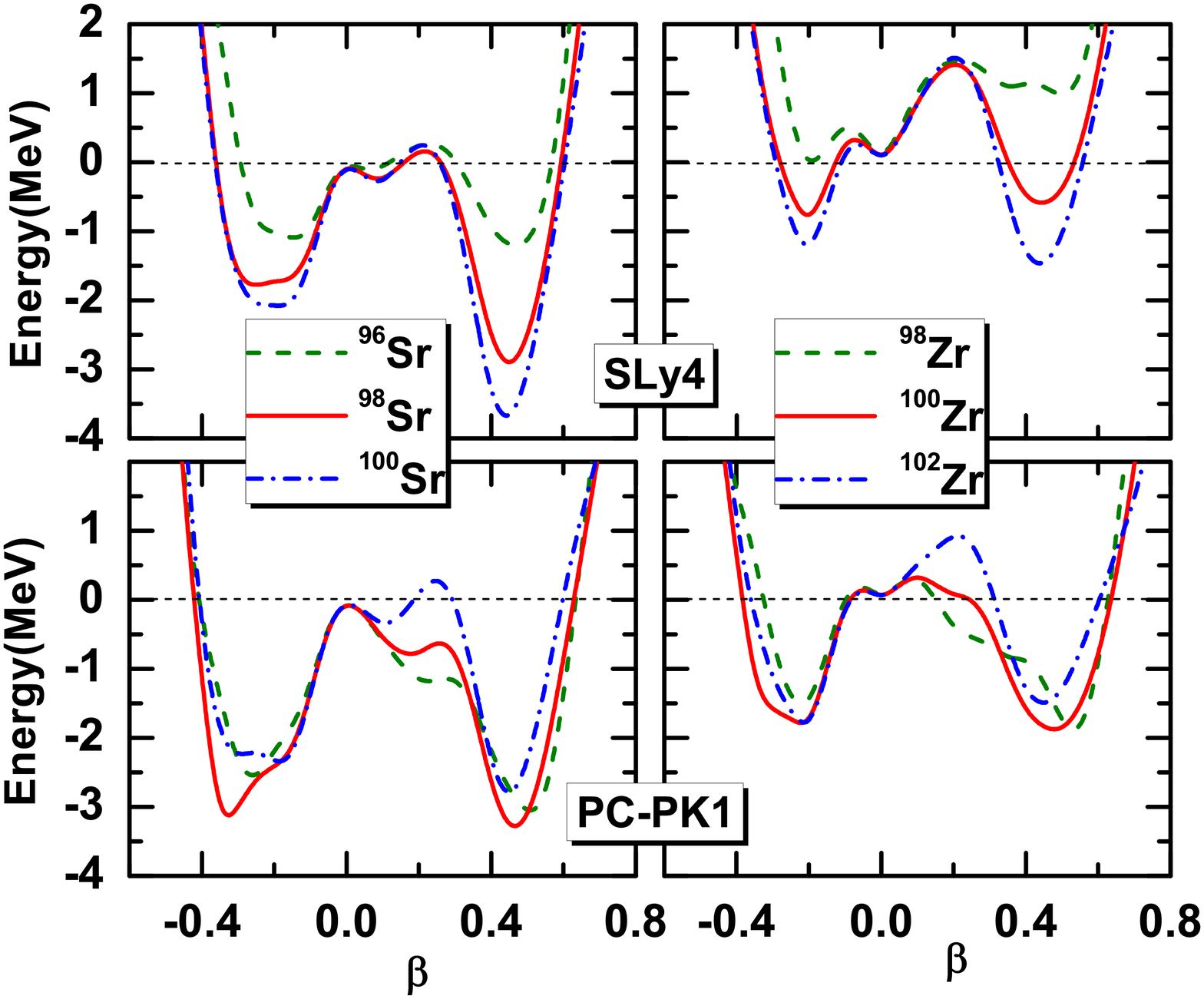}
 \caption{(Color online) The total energy in $^{96,98,100}$Sr and $^{98,100,102}$Zr as a function
 of axial deformation parameter $\beta$ from the constrained mean-field calculations with both the SLy4 and the PC-PK1 forces. All energies are normalized to the value at $\beta=0$. }
 \label{axial_energy}
\end{figure}

 To understand the evolution character of spectroscopic quantities around $N=60$ shown in Fig.~\ref{SLy4:be2}, we plot the deformation energy surfaces of $^{96,98,100}$Sr and $^{98,100,102}$Zr isotopes in $\beta$-$\gamma$ plane from the constrained mean-field calculations with both SLy4 and PC-PK1 forces in Figs.~\ref{Zr_Sr_SHF} and~\ref{Zr_Sr_RMF} respectively. In both calculations, there is always a weakly oblate deformed minimum coexisting with a prolate minimum in the deformation energy surfaces of $^{96,98,100}$Sr and $^{98,100,102}$Zr. Furthermore, the absolute minimum in both calculations is shifted from the oblate to the prolate side as the neutron number increases from $N=58$ to $N=60$.  However, the subtle balance between these two minima is quite different in these two calculations, as shown in Fig.~\ref{axial_energy}, where the total energy in $^{96,98,100}$Sr and $^{98,100,102}$Zr as a function of axial deformation parameter $\beta$ is plotted. The evolution of deformation energy curves, i.e., from weakly oblate deformed $^{96}$Sr, $^{98}$Zr to large prolate deformed $^{98}$Sr, $^{100}$Zr in SLy4 calculations, is much more rapid than that in PC-PK1 calculations. Moreover, the oblate and prolate minima in the deformation energy curves of $^{96,98,100}$Sr and $^{98,100,102}$Zr are quite close in energy. A very large mixing of oblate and prolate configurations is expected in their ground states. In contrary to the PC-PK1 calculations, the SLy4 force gives a broader or deeper oblate minimum in $^{96}$Sr and $^{98}$Zr respectively and deeper or broader prolate minimum in $^{98,100}$Sr and $^{100,102}$Zr respectively. In other words, the $^{96}$Sr and $^{98}$Zr are more oblate, while $^{98,100}$Sr and $^{100,102}$Zr are more prolate in the SLy4 calculations.

\begin{figure}[htp]
  \centering
 \includegraphics[width=8.2cm]{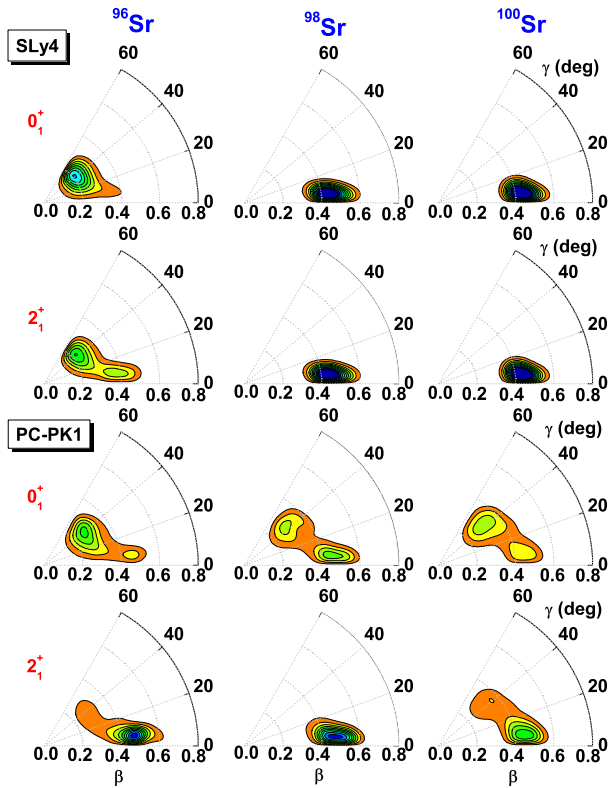}
 \caption{(Color online) The distribution of squared collective wave functions in the $\beta$-$\gamma$ plane for the $0^+_1$ and $2^+_1$ states in $^{96,98,100}$Sr from the 5DCH calculations with both the SLy4 and PC-PK1 forces.}
 \label{wfs}
\end{figure}

 The structural change in low-lying states is illustrated clearly in Fig.~\ref{wfs}, where the distribution of the squared collective wave functions $\rho_{I\alpha}$ in the $\beta$-$\gamma$ plane for the $0^+_1$ and $2^+_1$ states from the 5DCH calculations with both the SLy4 and the PC-PK1 forces is plotted. The $\rho_{I\alpha}$ is defined as
 \begin{equation}
  \rho_{I\alpha} (\beta,\gamma)
  = \sum_{K}\vert \Psi^{I}_{\alpha,K}(\beta,\gamma)\vert^2 \beta^3 \vert
  \sin3\gamma\vert,
 \end{equation}
 which follows the normalization condition,
 \begin{equation}
  \int^\infty_0 \beta d\beta \int^{2\pi}_0 d\gamma \rho_{I\alpha}(\beta,\gamma) = 1.
 \end{equation}
 Here, $\Psi^{I}_{\alpha,K}(\beta,\gamma)$ is the collective wave function that corresponds to
 the solution of 5DCH. The distributions of the squared collective wave functions in these two calculations are quite different. The SLy4 predicts a sudden transition from oblate $^{96}$Sr to very good prolate $^{98,100}$Sr, while, the PC-PK1 gives a coexistence picture for the ground states of $^{96,98,100}$Sr, with the dominant component changing from oblate to prolate and back to oblate again moderately. For the $2^+_1$ state in the calculations with the SLy4 force, the dominant component in $^{96}$Sr is oblate and it becomes well prolate in $^{98}$Sr. While in the calculations with the PC-PK1 force, the dominant component is already prolate in $^{96}$Sr.
 This picture provides a interpretation for the evolution character of $B(E2:2^+_1\rightarrow 0^+_1)$, $E_x(2^+_1)$ and $R_{4/2}$ shown in Fig.~\ref{SLy4:be2}. Compared with the experimental data for $B(E2:2^+_1\rightarrow 0^+_1)$ and $R_{4/2}$, the SLy4 (or PC-PK1) force overestimates (or underestimates) somewhat the slop of shape transition around $N=60$.

 \subsection{Isotope shifts and monopole transition strengths}

 Nuclear charge radii or isotopic shifts are good indicators of shape changes along isotopic chains. In Refs.~\cite{Guzman10} and \cite{Xiang12}, the evolution of charge radii with the neutron number in Sr and Zr isotopes around $N=60$ has been studied with the self-consistent constrained mean-field calculations. The charge radii corresponding to different local minima in the deformation energy surface are compared with the data. It has been shown that a rapid change in nuclear shape is essential to reproduce the experimental charge radii. However, in these two studies, the beyond mean-field correlation effect on
 nuclear charge radii has not been examined. In Ref.~\cite{Bender06}, the dynamic quadrupole correlation effect on charge radii of a large set of even-even nuclei has been studied in the framework of configuration mixing of projected axially deformed states based on a topological Gaussian overlap approximation. It has been shown that the dynamical correlation leads to an overall increase of radii, and it might also reduce the charge radii for some specific nuclei.
 Therefore, it is interesting to show the evolution of charge radii with the correlation effect from quadrupole fluctuation.
\begin{figure}[]
  \centering
 \includegraphics[width=8.2cm]{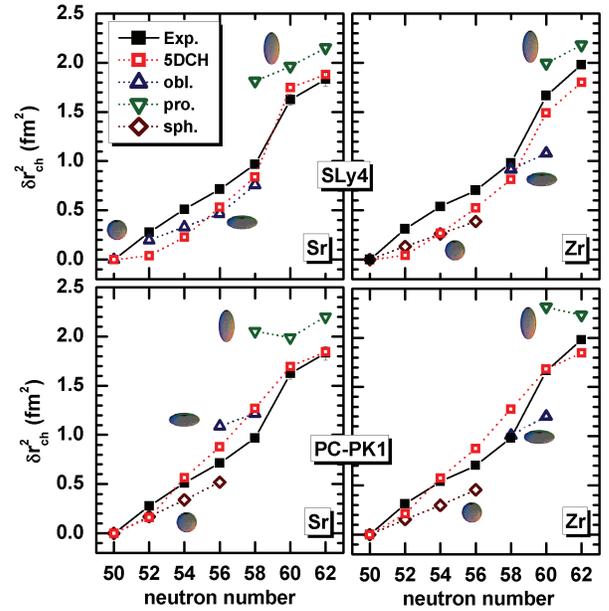}
 \caption{(Color online) The isotope shift (normalized to $N=50$) in Sr  and Zr isotopes as a function of neutron number from the mean-field calculations of both SHF+BCS with the SLy4 force and RMF+BCS with the PC-PK1 force. The corresponding 5DCH calculated results and experimental data from Refs.~\cite{Buchinger90,Campbel02} are given as well. The ``obl.", ``pro.", ``sph." are used to label the character of mean-field configurations.}
 \label{Charge_radii}
\end{figure}

 Figure~\ref{Charge_radii} displays the isotope shifts corresponding to different configurations in Sr and Zr isotopes from the mean-field calculations with both SLy4 and PC-PK1 forces. In addition, we present the 5DCH predicted isotope shifts, in which, the quadrupole fluctuation effect with triaxiality is included. It is shown clearly that the overall of isotope shifts could be reproduced much better by the 5DCH calculations with dynamic correlation effect.
 The shape evolution picture consistent with that exhibited in the $B(E2:2^+_1\rightarrow0^+_1)$, $R_{4/2}$ as well as the distribution of squared wave functions is shown again in the isotope shifts, a sudden rising of which from $N=58$ to $N=60$ due to the transition from weakly oblate deformed to large prolate deformed shapes is slightly overestimated (or underestimated) in the 5DCH calculations with the SLy4 (or PC-PK1) force.

\begin{figure}[]
  \centering
 \includegraphics[width=8.2cm]{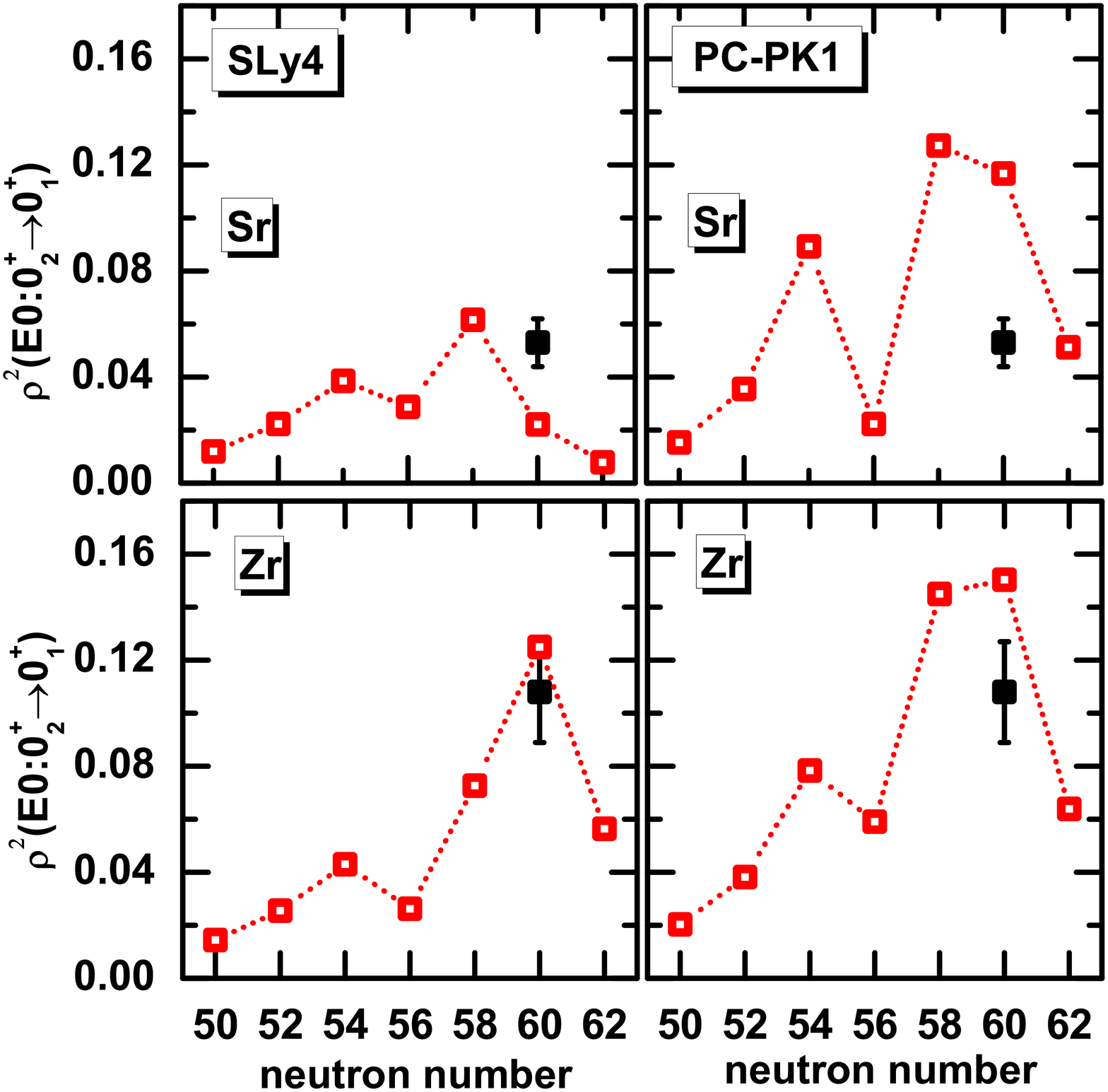}
 \caption{(Color online) The electric monopole transition strength $\rho^2(E0:0^+_2\rightarrow 0^+_1)$ from the 5DCH calculations with both the SLy4 and PC-PK1 forces, in comparison with the experimental data~\cite{Kibedi05,LBL}.}
 \label{E0}
\end{figure}

  Electric monopole transition strengths are a model-independent signature of the mixing of configurations with different mean-square charge radii~\cite{Heyde11} and therefore provide a good way to illustrate the occurrence of shape-coexistence. Figure~\ref{E0} shows the evolution of electric monopole transition strength $\rho^2(E0:0^+_2 \to 0^+_1)$, given by the off-diagonal element of charge radius,
\begin{equation}
 \rho^2(E0;{0^+_2 \to 0^+_1})
 =\left\vert \frac{\langle0^+_2\vert \sum_k e_kr_k^2 \vert 0^+_1\rangle}
  {eR^2_0}\right\vert^2,
\end{equation}
with $R_0=1.2A^{1/3}$~fm, with respect to the neutron number in neutron-rich Sr and Zr isotopes. The systematic of $\rho^2(E0:0^+_2\rightarrow 0^+_1)$ is similar in the calculations with the SLy4 and PC-PK1 forces. Both calculations predict the same peak position, i.e., at $^{96}$Sr and $^{100}$Zr, which are the nuclei before and after the dramatic transition respectively. Quantitatively, however, as expected from the distribution of squared wave functions for $^{96,98,100}$Sr in Fig.~\ref{wfs}, the SLy4 (or PC-PK1) force predicts a much weaker (or stronger) mixing of oblate and prolate configurations in their ground states, and therefore smaller (larger) $\rho^2(E0:0^+_2\rightarrow 0^+_1)$ values in $^{96,98,100}$Sr. Compared with the experiment data for $^{98}$Sr, the SLy4 (or PC-PK1) force underestimates (or overestimates) the $\rho^2(E0:0^+_2\rightarrow 0^+_1)$ value by a factor of two approximately. Similar evolution behavior is also shown in Zr isotopes. However, except for $^{98}$Zr, the difference in magnitude between the results of SLy4 and PC-PK1 calculations is smaller than that in Sr isotopes.

 \subsection{Single-particle energy levels and spin-orbit splitting}

\begin{figure*}[htp]
  \centering
 \includegraphics[width=8.2cm]{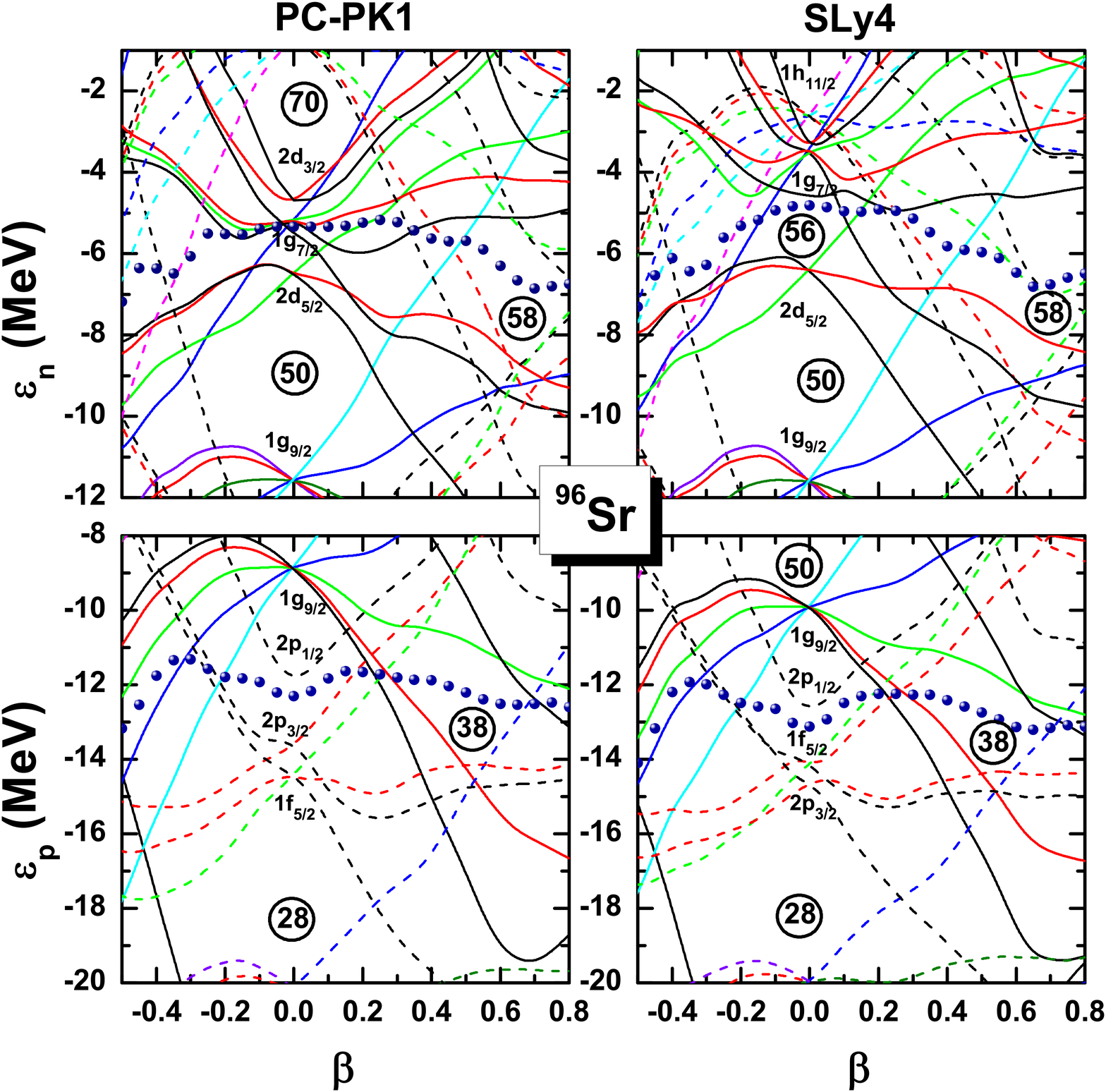}
 \includegraphics[width=8.2cm]{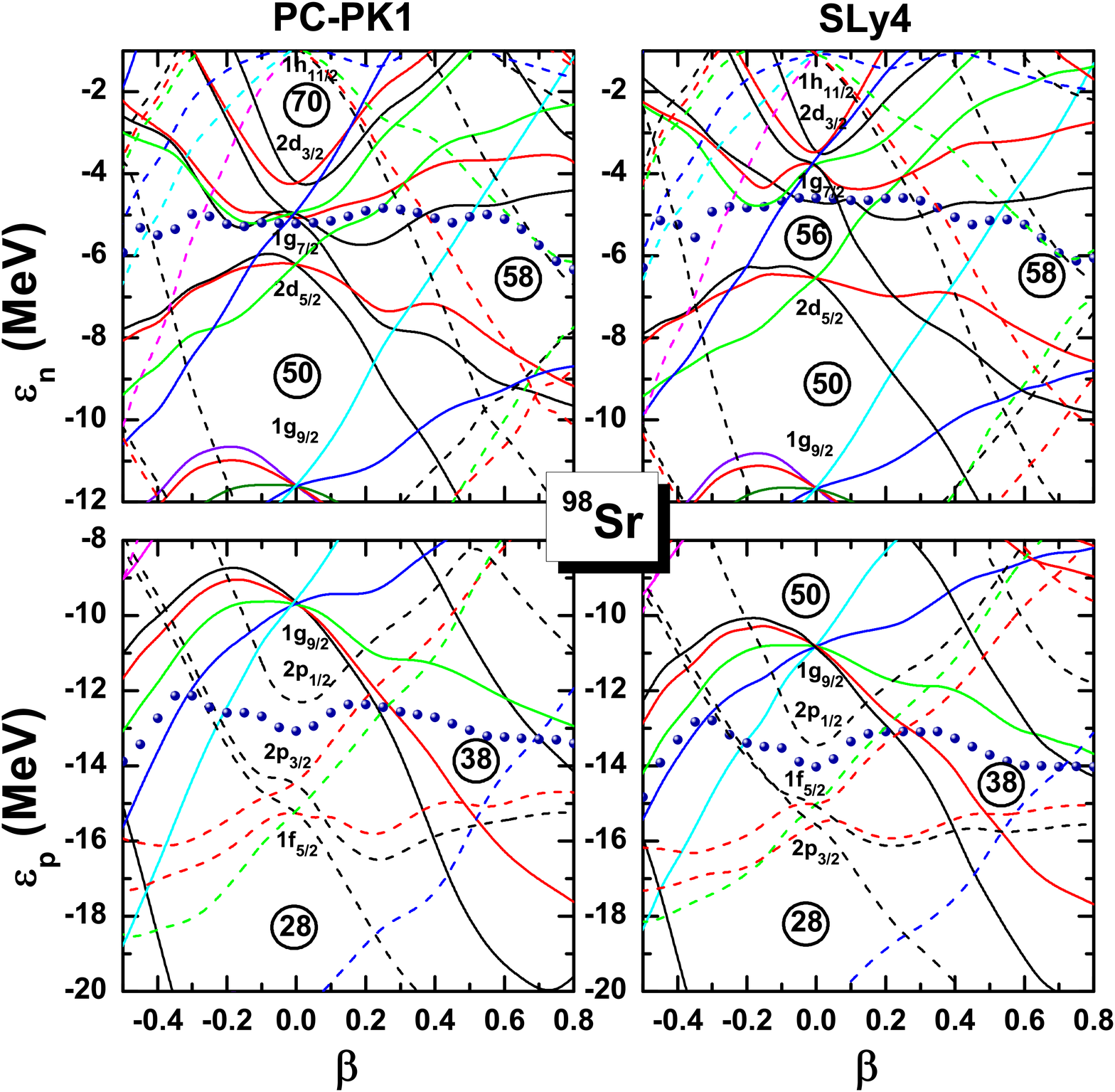}
 \caption{(Color online) Neutron and proton single-particle energy levels in (left panel) $^{96}$Sr and (right panel) $^{98}$Sr from the constrained mean-field calculations with both SLy4 and PC-PK1 forces. The dots denote the corresponding Fermi energy levels.}
 \label{eps-sr98}
\end{figure*}

 The nuclear low-lying states reflect information about the underlying single-particle structure and vice versa. Therefore, to understand the evolution character of collectivity in Sr and Zr isotopes around $N=60$ in the calculations with both the SLy4 and PC-PK1 forces, in Fig.~\ref{eps-sr98}, we plot the neutron and proton single-particle energy levels in $^{96,98}$Sr as functions of the axial deformation parameter $\beta$. In general, the minima in the deformation energy curve are associated with a shell effect due to the low level density around the Fermi energy. 
 It is shown in Fig.~\ref{eps-sr98} that, for the proton single-particle energy levels, both calculations predict a large shell gaps around the Fermi energy level in the prolate side, which is mainly formed by two levels split from the degenerate $\pi 1g_{9/2}$ state due to the deformation (or Jahn-Teller) effect. The size of this energy gap does not change too much from $^{96}$Sr to $^{98}$Sr in both calculations. It means that the proton plays a minor role in the rapid nuclear shape transition from the mean-field point of view. For the neutron single-particle energy levels, both calculations predict two evident shell gaps around the Fermi energy level in oblate and prolate sides, however, the details of the single-particle structure are quite different.

\begin{figure}[]
  \centering
 \includegraphics[width=8cm]{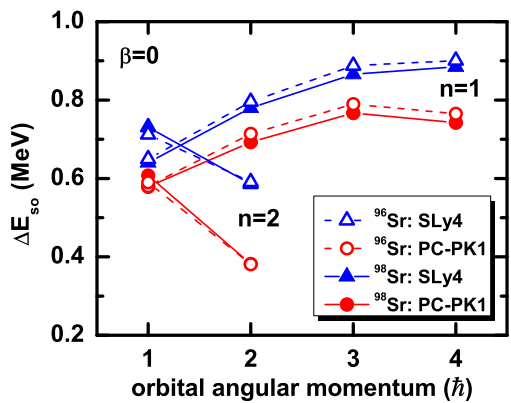}
 \caption{(Color online) Splitting of neutron spin-orbit doublet states [c.f.(\ref{sosplit})] in the spherical states of $^{96}$Sr (open symbols) and $^{98}$Sr (filled symbols) as a function of the orbital angular momentum $\ell$ from the mean-field calculations with both the SLy4 (triangle) and the PC-PK1 (circle) forces. }
 \label{splitting}
\end{figure}

 Compared with the PC-PK1 force, the SLy4 force predicts a stronger spin-orbit splitting for neutrons (by a factor of $\sim1.1$) for all the states, as shown in Fig.~\ref{splitting}, where the splitting of neutron spin-orbit doublet states,
\begin{equation}
\label{sosplit}
\Delta E_{\rm so}
 =\frac{\epsilon_{nlj_<}-\epsilon_{nlj_>}}{2\ell+1},~~j_{\gtrless}=\ell\pm1/2,
\end{equation}
 in the spherical states of $^{96}$Sr and $^{98}$Sr as a function of the orbital angular momentum $\ell$ is plotted. $\epsilon_{nlj_<}$ is the energy of single-particle state with quantum numbers $(n, \ell, j_<)$. Consequently, the position of $\nu 1g_{7/2}$ state is pushed up and the $\nu 1h_{11/2}$ state is pulled down compared with the PC-PK1 results as shown in Fig.~\ref{eps-sr98}. As a result, the shell gaps around the Fermi energy at the spherical point and the minima of the deformation energy curves are quite different. For the spherical point ($\beta=0$), a relatively large shell gap at $N=56$, which provides a mechanism responsible for the observed much higher $E_x(2^+_1)$ in $^{96}$Zr, is shown in the SLy4 calculation, but not in the PC-PK1 calculations.
 Moreover, compared with the PC-PK1 results for $^{96}$Sr, the position of $\nu 2d_{5/2}$ state is almost the same, but the $\nu 1h_{11/2}$ state is much lower in the SLy4 results. As a result, the shell gap around the Fermi level in the prolate side, mainly formed by the $K=3/2$ component of $\nu 2d_{5/2}$ orbit, the intruded $K=1/2$ component of $\nu 2f_{7/2}$ orbit, and other two levels with $K=3/2, 5/2$ split from $\nu 1h_{11/2}$ orbit, is much smaller than that in the PC-PK1 calculations. In $^{98}$Sr, the energy of $\nu 1h_{11/2}$ state is shifted up, which broadens the shell gap significantly in the prolate side. This big change in the energy gap around Fermi energy is responsible for the sudden onset of large prolate deformation at $N=60$ given by the SLy4 calculations. In the PC-PK1 calculations, however, this change in the shell gap of prolate side is more moderate. We note that, similar as the PC-PK1 results, the shift of $\nu 1h_{11/2}$ state in $^{96,98}$Sr from the Gogny D1S calculations is small and the change in the shell gap of prolate side is not significant~\cite{CEA}. On the other hand, compared with the SLy4 force, the PC-PK1 predicts a larger shell gaps in the oblate side of $^{96,98}$Sr, which provides the mechanism responsible for the strong mixing of prolate and oblate shapes in their ground states.

 \section{Summary}
 \label{Sec.IV}
In summary, the rapid structural change in low-lying collective excitation states of neutron-rich Sr and Zr isotopes has been studied by solving a 5DCH with parameters determined from both the RMF and SHF calculations. Pair correlations are treated in the BCS method with either a separable pairing force or a density-dependent zero-range force. The isotope shifts, excitation energies, electric monopole and quadrupole transition strengths have been calculated and compared with corresponding experimental data. The calculated results with both the PC-PK1 and the SLy4 forces exhibit a picture of spherical-oblate-prolate shape transition in neutron-rich Sr and Zr isotopes. However, compared with the experimental data, the PC-PK1 (or SLy4) force predicts a more moderate (or dramatic) change in most of the collective properties around $N=60$ and a much stronger (or weaker) mixing between oblate and prolate configurations in their ground states. The difference between these two calculations is mainly because of the quite different structure in neutron single-particle states, mostly caused by the different spin-orbit interaction strengths. Moreover, the sudden broadening of neutron shell gap in the prolate side, mainly formed by the $K=3/2$ component of $\nu 2d_{5/2}$, the intruded $K=1/2$ component of $\nu 2f_{7/2}$, and other two components of $\nu 1h_{11/2}$ state, has been shown to be responsible for the rapid shape transition at $N=60$. However, it has to be pointed out that the rapid change in the excitation energy of the first $2^+$ state has not been reproduced in the calculations with both the PC-PK1 and SLy4 forces. In particular, even though the SHF+BCS calculation with the SLy4 force indeed predicts a sizable spherical shell gap at $N=56$, the corresponding 5DCH calculation is not able to reproduce the suddenly increased excitation energy for the first $2^+$ state at $^{96}$Zr. A further beyond mean-field investigation is required.

\begin{acknowledgments} 
JMY gratefully acknowledges a postdoctoral fellowship from the
F.R.S.-FNRS (Belgium) and fruitful discussions with Paul-Henri Heenen. This work was partly supported by the Major State Basic Research Developing Program 2007 CB815000, the National Science Foundation of China under Grants No. 11105111,  No. 11105110, No. 11175002, No. 10975008 and No. 10947013, the Fundamental Research Funds for the Central Universities (XDJK2010B007 and XDJK2011B002), the Research Fund for the Doctoral Program of Higher Education under Grant No. 20110001110087 and the Southwest University Initial Research Foundation Grant to Doctor (SWU109011 and SWU110039).
\end{acknowledgments}


\begin{thebibliography}{90}%


\bibitem{Jung80}     G. Jung \emph{et al.}, Phys. Rev. \textbf{C22}, 252 (1980).
\bibitem{Schussler80}F. Schussler, J, A. Pinston, B. Monnand and A. Moussa,
                     Nucl. Phys. \textbf{A339}, 415 (1980).
\bibitem{Kawade82}   K. Kawade \emph{et al.}, Z. Phys. \textbf{A304}, 293 (1982).
 \bibitem{Eberth88}  J. Eberth and K. Sistemich (Eds.),
                     Nuclear Structure of the Zirconium Region, Proceedings of the International Workshop (Springer-Verlag, 1988).
\bibitem{Mach89}     H. Mach \emph{et al.}, Phys. Lett.  \textbf{B230}, 21 (1989).
\bibitem{Mach91}     H. Mach \emph{et al.}, Nucl. Phys. \textbf{A523}, 197 (1991).
\bibitem{Lhersonneau94}G. Lhersonneau \emph{et al.},
                     Phys. Rev. \textbf{C49}, 1379 (1994).
\bibitem{Urban01}    W. Urban \emph{et al.},  Nucl. Phys. \textbf{A689}, 605 (2001).
\bibitem{Hager06}    U. Hager \emph{et al.}, Phys. Rev. Lett. \textbf{96}, 042504 (2006).

\bibitem{Goodin07}   C. Goodin \emph{et al.}, Nucl. Phys. \textbf{A787}, 231 (2007).
\bibitem{Charlwood09}F. C. Charlwood \emph{et al.}, Phys. Lett. \textbf{B674},  23 (2009).


 \bibitem{Federman78} P. Federman and S. Pittel,
                      Phys. Lett. \textbf{B77}, 29 (1978);
                      Phys. Rev. \textbf{C20}, 820 (1979).
\bibitem{Kumar85}  A. Kumar and M. R. Gunye, Phys. Rev. \textbf{C32},  2116 (1985).
\bibitem{Galeriu86}D. Galeriu, D. Bucurescu, and M. Ivaqcu,
                   J. Phys. G: Nucl. Part. Phys. \textbf{12}, 329 (1986).
\bibitem{Michiaki90}S. Michiaki and A. Akito, Nucl. Phys. \textbf{A515}, 77 (1990).
\bibitem{Moller95}  P. M\"{o}ler, J. R. Nix, W. D. Myers, and W. J. Swiatecki,
                    At. Data Nucl. Data Tables \textbf{59}, 185 (1995).
\bibitem{Skalski97} J. Skalski, S. Mizutori, and W. Nazarewicz,
                    Nucl. Phys. \textbf{A617}, 282 (1997).

\bibitem{Lalazissis99} G. A. Lalazissis, S. Raman, P. Ring,
                       At. Data Nucl. Data Tables \textbf{71}, 1 (1999).
 \bibitem{Holt00}   A. Holt, T. Engeland, M. Hjorth-Jensen, and E. Osnes,
                    Phys. Rev. \textbf{C61}, 064318 (2000)
\bibitem{Xu02}        F. R. Xu, P.M. Walker, and R. Wyss, Phys. Rev. \textbf{C65}, 021303(R) (2002).
\bibitem{Ramos05}  J. Garc\'{i}a-Ramos, K. Heyde, R. Fossion, V. Hellemans, and S. De Baerdemacker,
                      Eur. Phys. J. \textbf{A26}, 221  (2005).
\bibitem{Geng05}      L. S. Geng, H. Toki, and J. Meng, Prog. Theor. Phys. {\bf113}, 785 (2005).
\bibitem{Lalkovski09} S. Lalkovski and P. Van Isacker. Phys. Rev. \textbf{C79}, 044307 (2009).
\bibitem{Isacker10}   P. Van Isacker, A. Bouldjedri and S. Zerguine. Nucl. Phys. \textbf{A836}, 225 (2010).
\bibitem{Boyukata10}  M. B\"{o}y\"{u}kata, P. Van Isacker and i. Uluer.
                      J. Phys. G: Nucl. Part. Phys. \textbf{37}, 105102 (2010).


\bibitem{Werner94} T. R. Werner, J. Dobaczewski, M. W. Guidry, W. Nazarewicz, and J. A. Sheikh,
                   Nucl. Phys. \textbf{A578}, 1 (1994).
\bibitem{Bonche85} P. Bonche, H. Flocard, P. H. Heenen, S. J. Krieger, M. S. Weiss,
                   Nucl. Phys. \textbf{A443}, 39 (1985).
\bibitem{Verma08}  S. Verma, P. A. Dar, and R. Devi, Phys. Rev. \textbf{C77}, 024308 (2008).
\bibitem{Sieja09} K. Sieja, F. Nowacki, K. Langanke, and G. Mart\'\i{}nez-Pinedo,
                  Phys. Rev. \textbf{C79}, 064310 (2009).

 \bibitem{Lalazissis04} G. A. Lalazissis, P. Ring, and D. Vretenar (Eds.), Extended Density Functionals in Nuclear Structure Physics, Lecture Notes in Physics 641, (Springer, Heidelberg 2004).

\bibitem{Lu11}  B.-N. Lu, E.-G. Zhao, and S.-G. Zhou, Phys. Rev. {\bf C84}, 014328 (2011).

\bibitem{Guzman10} R. Rodr\'{i}guez-Guzm\'{a}n, P. Sarriguren, L. M. Robledo and S. Perez-Martin,
                   Phys. Lett. \textbf{B691}, 202 (2010).
 \bibitem{Xiang12}    J. Xiang, Z.P. Li, Z.X. Li, J.M. Yao, and J. Meng,
                      Nucl. Phys. {\textbf A873}, 1 (2012).
 \bibitem{Skalski93}   J. Skalski, P.-H. Heenen, and P. Bonche,
                       Nucl. Phys. \textbf{A559}, 221 (1993).

 \bibitem{Valor00}  A. Valor, P. H. Heenen, and P. Bonche, Nucl. Phys. {\bf A671}, 145 (2000).
 \bibitem{Guzman02} R. Rodr\'{\i}guez-Guzm\'{a}n, J. L. Egido, and L. M. Robledo,
                    Nucl. Phys. {\bf A709}, 201 (2002).
 \bibitem{Niksic06} T. Nik\v{s}i\'{c}, D. Vretenar, and P. Ring,
                    Phys. Rev. {\bf C73}, 034308 (2006); {\bf C74}, 064309 (2006).

 \bibitem{Bender08}   M. Bender and P.-H. Heenen, Phys. Rev. {\bf C78}, 024309 (2008).
 \bibitem{Yao09}      J. M. Yao, J. Meng, P. Ring, and D. Pena Arteaga,
                      Phys. Rev. {\bf C79}, 044312 (2009).
 \bibitem{Yao10}      J. M. Yao, J. Meng, P. Ring and D. Vretenar,
                       Phys. Rev. {\bf C81}, 044311 (2010).
 \bibitem{Rodriguez10} T. R. Rodr\'{\i}guez and J. L. Egido,
                       Phys. Rev. {\bf C81}, 064323 (2010).

 \bibitem{Yao11-Mg}    J. M. Yao, H. Mei, H. Chen {\em et al}.,
                       Phys. Rev. {\bf C83}, 014308 (2011).
 \bibitem{Yao11-C}     J. M. Yao, J. Meng, P. Ring {\em et al}.,
                       Phys. Rev. {\bf C84}, 024306 (2011).

 \bibitem{Tomas11PRC}  T. R. Rodr\'{\i}guez and J. L. Egido, Phys. Rev. {\bf C84}, 051307 (2011)
 \bibitem{Tomas11PLB}  T. R. Rodr\'{\i}guez and J. L. Egido, Phys. Lett. {\bf B705},  255 (2011).


 \bibitem{Libert99}    J. Libert, M. Girod, and J.-P. Delaroche, Phys. Rev. {\bf C60}, 054301 (1999).
 \bibitem{Prochniak04} L. Pr\'{o}chniak, P. Quentin, D. Samsoen, and J. Libert,
                       Nucl. Phys. {\bf A730}, 59 (2004).
 \bibitem{Niksic09}    T. Nik\v{s}i\'{c}, Z. P. Li, D. Vretenar {\em et al}.,
                       Phys. Rev. {\bf C79}, 034303 (2009).
 \bibitem{Li09}        Z. P. Li, T. Nik\v{s}i\'{c}, D. Vretenar {\em et al}.,
                       Phys. Rev. {\bf C79}, 054301 (2009).

 \bibitem{Niksic11}    T. Nik\v{s}i\'{c}, D. Vretenar and P. Ring,
                       Prog. Part. Nucl. Phys. {\bf 66}, 519 (2011).

 \bibitem{Yao11-lambda} J. M. Yao, Z. P. Li, K. Hagino {\em et al}.,
                       Nucl. Phys. {\textbf A868-869}, 12 (2011).

 \bibitem{Delaroche10} J. -P. Delaroche, M. Girod, J. Libert {\em et al}.,
                       Phys. Rev. {\textbf C81}, 014303 (2010).

 \bibitem{Pro.99}  L. Pr\'{o}chniak, K. Zajac, K. Pomorski {\em et al}.,
                   Nucl. Phys. {\bf A648}, 181 (1999).

\bibitem{Zhao10} P. W. Zhao, Z. P. Li, J. M. Yao, and J. Meng,
                 Phys. Rev. \textbf{C82}, 054319  (2010).

\bibitem{Tian09}   Y. Tian and Z. Y. Ma, and P. Ring,
                   Phys. Lett. \textbf{B676}, 44 (2009).
\bibitem{Niksic10}  T. Nik\v{s}i\'{c}, P. Ring, D. Vretenar, Y. Tian, and Z. Y. Ma,
                     Phys. Rev. \textbf{C81}, 054318 (2010).
 \bibitem{Chabanat98} E. Chabanat, P. Bonche, P. Haensel, J. Meyer, and R. Schaeffer,
                      Nucl. Phys. {\bf A635},  231 (1998); {\bf643}, 441(E) (1998).

\bibitem{Rig99}      C. Rigollet, P. Bonche, H. Flocard, P.-H. Heenen,
                      Phys. Rev. \textbf{C59}, 3120 (1999).

 \bibitem{Ing.56}      D. R. Inglis, Phys. Rev. {\bf 103}, 1786 (1956).
 \bibitem{Bel.61}      S. T. Belyaev, Nucl. Phys. {\bf 24}, 322 (1961).
 \bibitem{GG.79}       M. Girod and B. Grammaticos, Nucl. Phys. {\bf A330}, 40 (1979).
 \bibitem{Bonche05}    P. Bonche, H. Flocard, and P.-H. Heenen,
                       Comput. Phys. Commun. {\bf 171}, 49 (2005).

 \bibitem{NNDC} National Nuclear Data Center, Brookhaven National Laboratory, http://www.nndc.bnl.gov/index.jsp.

 \bibitem{Bender06} M. Bender, G. F. Bertsch, and P.-H. Heenen,
                    Phys. Rev. {\textbf C73}, 034322 (2006).

 \bibitem{Buchinger90} F. Buchinger, E. B. Ramsay, E. Arnold {\em et al}.,
                       Phys. Rev. {\textbf C41}, 2883 (1990).

  \bibitem{Campbel02}  P. Campbell, H. L. Thayer, J. Billowes  {\em et al}.,
                       Phys. Rev. Lett. {\textbf 89}, 082501 (2002)

\bibitem{Kibedi05}     T. Kib\'{e}di and R. H. Spear,
                       At. Data Nucl. Data Tables \textbf{89}, 77 (2005).

\bibitem{LBL}          http://ie.lbl.gov/TOI2003/GammaSearch.asp.



  \bibitem{Heyde11} K. Heyde and J. L. Wood, Rev. Mod. Phys. {\bf 83}, 1467 (2011).

  \bibitem{CEA} http://www-phynu.cea.fr/science\_en\_ligne/carte\_potentiels
               \_microscopiques/carte\_potentiel\_nucleaire.htm.



 \end{thebibliography}
\end{document}